\documentclass[%
 reprint,
superscriptaddress,
 amsmath,amssymb,
 aps,
]{revtex4-1}

\usepackage{graphicx}
\usepackage{dcolumn}
\usepackage{bm}
\usepackage{hyperref}

\begin{document}

\preprint{APS/123-QED}

\title{Destroying the Event Horizon of a Rotating Black-Bounce Black Hole}

\author{Lai Zhao}

\author{Zhaoyi Xu}%
\email{zyxu@gzu.edu.cn(Corresponding author)}
\affiliation{%
 College of Physics,Guizhou University,Guiyang,550025,China
}%

\date{\today}

\begin{abstract}
For a rotating black hole to be nonsingular, it means that there are no spacetime singularities at its center. The destruction of the event horizon of such a rotating black hole is not constrained by the weak cosmic censorship conjecture, which may provide possibilities to understand the internal structure of black hole event horizons. In this paper, we employ test particles with large angular momentum and a scalar field with large angular momentum to investigate the potential of destroying the event horizon of rotating Black-Bounce black holes. Additionally, we investigate the possibility of destroying the event horizon of a rotating Black-Bounce black hole by considering test particles with large angular momentum and scalar fields with large angular momentum, covering the entire range of the rotating Black-Bounce black hole. We analyze the influence of the parameter $m$ on the possibility of destroying the event horizon in this spacetime. Our analysis reveals that under extreme or near-extreme conditions, the event horizon of this spacetime can potentially be destroyed after the absorption of particles energy and angular momentum, as well as the scattering of scalar fields. Additionally, we find that as the parameter $m$ increases, the event horizon of this spacetime model becomes more susceptible to destruction after the injection of test particles or the scattering of scalar fields.
\begin{description}
\item[Keywords]
Scalar field; Rotating black-bounce black hole;  Weak cosmic censorship conjecture; Spacetime singularity.
\end{description}
\end{abstract}

\maketitle


\section{\label{sec:level1}Introduction}
The importance of the singularity theorem has been demonstrated by scholars such as Penrose and Hawking\cite{PhysRevLett.14.57,1970RSPSA.314..529H,hawking_ellis_1973}. This theorem proves that under certain conditions of matter-energy and initial conditions, gravitational collapse leads to the formation of spacetime singularities\cite{Wald1999}. This implies that in some cases, gravitational collapse can cause the discontinuity and unpredictability of the geometric structure of spacetime, in other words, the physical laws near the singularity are threatened. In order to make physical laws more universal, physicists have attempted to explain the behavior near singularities through other theories, such as string theory or loop quantum gravity\cite{PhysRevLett.96.141301,BRANDENBERGER1989391}. These theories aim to provide new interpretations or avoid the appearance of singularities by quantizing gravity and considering factors such as higher dimensions and microscopic structures. Some studies suggest that certain quantum corrections may suppress the formation of singularities and propose models that could potentially address the problem of black hole singularities\cite{Steinhardt_2002,PhysRevLett.90.151301,Yang2022DestroyingTE}. If naked singularities exist in spacetime, even in regions far from the singularity, where causally connected events are present, physical laws cannot be predicted using general relativity. In order to ensure the validity of general relativity in regions far from naked singularities, the British physicist Roger Penrose proposed the hypothesis of the weak cosmic censorship conjecture\cite{article}.

This hypothesis suggests that the gravitational field of a black hole would hide any singularities within its event horizon, thereby avoiding their impact on the external world. This restoration of predictability in general relativity allows observers outside the black hole to make reliable predictions. According to the cosmic censorship conjecture, the gravitational field of a black hole would “cloak” any singularities, preventing their observation by the external world. Consequently, if singularities exist, they should be part of the interior of the black hole and not affect the external region. Although the weak cosmic censorship conjecture remains a hypothesis, it has been confirmed in various aspects, such as numerical simulations studying the collapse of dust clouds or matter\cite{Christodoulou1984,Mizuno2016,PhysRevD.38.2972,1975PhRvD..12.3072E,Song_2021}, or the simulation of black hole collisions in higher-dimensional spaces\cite{Andrade2018CosmicCV,Andrade2019,PhysRevLett.103.131102}.

As of now, the cosmic censorship conjecture remains a hypothesis and cannot be described by a rigorous mathematical formulation\cite{Wald1999}, requiring further experimental verification and theoretical exploration. However, the weak cosmic censorship conjecture has become an important pillar in the study of black hole theory. In recent years, there have been new developments in the research of the cosmic censorship conjecture\cite{gao2022blackholes}, including optimization of the idea of weak cosmic censorship, weak cosmic censorship in higher-dimensional asymptotically flat spacetimes, and its extension to weak cosmic censorship in asymptotically anti-de Sitter spacetimes\cite{ZENG2020115162,Hong_2019}.There are two commonly used methods to verify the weak cosmic censorship conjecture. One is through numerical simulations, such as studying the evolution of collapsing matter or the collision of two black holes\cite{Christodoulou1984,PhysRevLett.66.994,PhysRevLett.70.9,PhysRevLett.59.2137,PhysRevLett.103.131102,Andrade2019}. The other is through thought experiments. A notable thought experiment proposed by Wald\cite{PhysRevD.100.044043,PhysRevD.101.084005,WALD1974548} involves large-angular-momentum test particles colliding with extreme or near-extreme black holes to investigate whether the event horizon can be destroyed to form a naked singularity. In a series of these thought experiments, a paper by Veronika E. Hubeny demonstrates that near-extreme Reissner-Nordström black holes can have their event horizons destroyed by carefully chosen particles, without considering backreaction effects\cite{PhysRevD.59.064013}. Many more thought experiments have been constructed, revealing behaviors that defy the weak cosmic censorship conjecture\cite{PhysRevD.87.044028,Jacobson2009OverspinningAB,PhysRevD.96.064010,Cardoso2015CosmicCA}. However, Wald and Sorce’s verification of the Kerr-Newman black hole found that under certain second-order perturbations, the weak cosmic censorship conjecture seems to be restored\cite{PhysRevD.96.104014}. Subsequent thought experiments and analyses on BTZ black holes or other types of black holes have further demonstrated that extreme black holes cannot be disrupted to form naked singularities\cite{Chen_2020,PhysRevD.83.104037,Semiz2011,Tóth2012}. In recent years, the weak cosmic censorship conjecture has been extended to loop quantum gravity, where the presence of quantum correction coefficients can destroy the event horizon, thus violating the weak cosmic censorship conjecture\cite{Yang2022DestroyingTE}.

In addition to using test particles to destroy the event horizon, another method employed to verify this hypothesis involves the scattering of scalar fields. When considering the scattering of classical scalar fields, it has been found that extreme and near-extreme black holes cannot have their event horizons destroyed by classical scalar fields\cite{Gwak2018WeakCC,LIANG2021115335,Gonçalves2020,Gwak_2020,CV}. However, considering the existence of super-radiance and quantum effects\cite{PhysRevLett.124.211101,brito2020superradiance, britannica2023hawkingradiation, jones2023blackholes}, black holes can absorb or emit energy from the scalar field, resulting in the phenomenon of black hole superradiance. When studying the scattering of quantum-corrected black holes with scalar fields, researchers have discovered that both extreme and near-extreme black holes can have their event horizons destroyed by scalar fields, thus violating the weak cosmic censorship conjecture. This also implies that when quantum effects are taken into account, certain near-extremal black holes, due to super-radiance, can absorb a dangerous amount of energy from a scalar field, leading to the destruction of their event horizons and the exposure of naked singularities\cite{PhysRevD.79.101502,PhysRevLett.100.121101,PhysRevLett.99.181301}.

The black-bounce spacetime is a type of deformable Schwarzschild spacetime, which was first proposed by Simpson and Visser\cite{Simpson_2019}. Their work attempted to establish a connection between black holes and wormholes by introducing a parameter. Since the introduction of the black-bounce spacetime, extensive research has been conducted on this topic. For example, Lobo et al. generalized the black-bounce spacetime to a more general form, enable it to be derived from the Einstein field equations, and also studied the dynamical properties of the black-bounce metric\cite{PhysRevD.101.124035,PhysRevD.103.084052}. Recently, in reference\cite{Xu2021}, Zhaoyi Xu et al. extended the black-bounce spacetime into a rotating black-bounce spacetime using the Newman-Janis algorithm (NJA).

In general, for a regular black hole, there is no singularity at its center, and it is instead replaced by a ring singularity or a transition surface. The destruction of its event horizon does not violate the weak cosmic censorship conjecture\cite{Yang2022DestroyingTE}. The disruption of the event horizon of a regular black hole was first proposed in reference\cite{PhysRevD.87.124022}. Their conclusion suggests that the event horizon of a regular black hole can be disrupted. The rotating black-bounce model\cite{Xu2021} describes a spacetime similar to a black hole. This black hole is a regular one, but what's interesting is that this regular black hole represents not just a black hole, but due to the presence of parameter $m^2$, the black hole can transition into a wormhole. In this process, the internal structure of the rotating black-bounce black hole may be quite fascinating. For rotating black-bounce model depicts a regular black hole that exhibits a bouncing phenomenon in its metric, thereby avoiding the issues of infinite density and singularities. Inspired by recent generalization of gedanken experiments by others, we also investigate the destruction of the event horizon by using test particles and scalar fields incident on the rotating Black-Bounce black hole, and discuss the impact of parameters on the destruction of the event horizon.

In this paper, In section 2,we provides an introduction to the rotating Black-Bounce black hole. In section 3 and section 4,we discuss the destruction of its event horizon by test particles and scalar fields, respectively, and explore the effects of parameters on the destruction of the event horizon. In the final section, we present some brief discussion and conclusion.
\section{\label{sec:level2}Rotating Black-Bounce black hole metric}
For the rotating Black-Bounce spacetime model, the metric takes the following form\cite{Xu2021}
\begin{equation}\label{1}
\begin{split}
ds^2=&-\left(1-\frac{2M\left(r^2+m^2\right)r^k}{\left(r^{2n}+m^{2n}\right)^\frac{k+1}{2n}\Sigma^2}\right)dt^2+\frac{\Sigma^2}{\Delta}dr^2\\
&-\frac{4a\sin^2{\theta}}{\Sigma^2}\frac{M\left(r^2+m^2\right)r^k}{\left(r^{2n}+m^{2n}\right)^\frac{k+1}{2n}}d\phi dt+\Sigma^2d{\theta}^2\\
&+\frac{\sin^2{\theta}}{\Sigma^2}((r^2+m^2+a^2)^2-a^2\Delta\sin^2{\theta})d{\phi}^2.
\end{split}
\end{equation}
The parameter $a$ in the above equation represents the black hole spin,where the metric functions are\\
\begin{equation}\label{2}
\begin{split}
a&=\frac{J}{M},\\
\Sigma^2&=r^2+m^2+a^2\cos^2{\theta},\\
\Delta&=h(r)f(r)+a^2\\&=r^2+m^2-\frac{2M(r^2+m^2)r^k}{(r^{2n}+m^{2n})^{\frac{k+1}{2n}}}+a^2.
\end{split}
\end{equation}
The parameter $m$ is a non-negative value, while $k$ , $n$ =1,2,3\ldots

For equation(\ref{1}), when the parameters $k = 1$, $n = 1$, and $a = 0$, the Kerr-like space-time metric degenerates into a Szekeres-Vřes spacetime. When $m = k = 0$ and $n = 1$, this spacetime becomes a Kerr black hole. If $m = k = 0$, $n = 1$, and $a = 0$, the spacetime metric becomes spherically symmetric. In the aforementioned process, the parameter $m$ is particularly interesting. As $m$ varies from zero to infinity, the spacetime metric transforms into Kerr black holes, Kerr-like black holes, single-channel rotating wormholes, and traversable wormholes.

Next, an analysis is conducted on the metric described by equation (\ref{1}), where the event horizon of this spacetime is determined by the equation $ \Delta= 0$, By utilizing equation (\ref{2}), the calculation yield
\begin{equation}\label{3}
 \frac{r_h}{\Gamma}=1\pm\sqrt{1-\frac{{a^2+m}^2}{\Gamma^2}},
\end{equation}
where
\begin{equation}\label{4}
\Gamma=\frac{M(r_h^2+m^2)r_h^{k-1}}{(r_h^{2n}+m^{2n})^{\frac{k+1}{2n}}}=M\beta,
\end{equation}
and $r_h$ represents the event horizon.

From the above equation for the event horizon (\ref{3}), it can be observed that $(a^2+m^2)\le\Gamma^2$ represents the existence of an event horizon in this spacetime metric. However, when the black hole spin parameter $a$ changes, i.e.,  $(a^2+m^2)>\Gamma^2$, the event horizon ceases to exist in this spacetime metric. This is particularly interesting as it indicates that certain changes in the spin parameter can lead to the destruction of the event horizon. The focus of this study will therefore be on the scenario where the event horizon exists.

We have calculated the event horizon of this spacetime,and next we will calculate the surface area of the event horizon and the angular velocity at the event horizon.

The surface area of the event horizon is
\begin{equation}\label{5}
A=4\pi(r_h^2+m^2+a^2),
\end{equation}
and the angular velocity at the event horizon is 
\begin{equation}\label{6}
\Omega_H=-\frac{g_{03}}{g_{33}}=\frac{a}{(r_h^2+m^2+a^2)}.
\end{equation}
\section{\label{sec:level3}Destruction of the black hole event horizon by test particles}
In Section \ref{sec:level2}, it can be seen that by changing the black hole spin parameter $a$, the event horizon in this spacetime metric can disappear. This can be observed from the event horizon formula (\ref{3})
\begin{equation}\label{7}
 \frac{r_h}{\Gamma}=1\pm\sqrt{1-\frac{{a^2+m}^2}{\Gamma^2}}.
\end{equation}
From the event horizon formula, it can be observed that when $(a^2+m^2)\le\Gamma^2$, the rotating black-bounce spacetime represents a black hole. However, when the condition $(a^2+m^2)>\Gamma^2$ is satisfied, resolving it yields
\begin{equation}\label{8}
J>M^2\sqrt{\beta^2-\frac{m^2}{M^2}},
\end{equation}
where $\Gamma=M\beta$, at this point, the event horizon of the spacetime disappears.

In this section, in order to destroy the event horizon of the rotating Black-Bounce black hole, it is sufficient to introduce particles or scalar fields with large angular momentum into this model. This causes a change in the black hole spin parameter, resulting in a composite system where the event horizon does not exist. Specifically, if the composite system satisfies the condition $ J'> {M'}^2\sqrt{\beta^2-\frac{m^2}{M^2}}$, the event horizon of this spacetime can be destroyed.

Under the spacetime of rotating black-bounce, the motion equation of a test particle with mass ($\mu $) can be expressed using the geodesic equations. In this process, adopting the affine parameter as the proper time ($\tau$) to describe the motion parameters of the test particle, the motion equation of this test particle can be written in the following form

\begin{equation}\label{9}
\frac{{d^2x^\mu}}{{d\tau^2}} + \Gamma^\mu_{\alpha\beta} \frac{{dx^\alpha}}{{d\tau}} \frac{{dx^\beta}}{{d\tau}} = 0.
\end{equation}
The equation above can be obtained through the Euler-Lagrangian equation of the test particle. The motion of the test particle can be described using the Lagrangian, which has the following expression
\begin{equation}\label{10}
L=\frac{1}{2}\mu g_{\mu\nu}\frac{dx^\mu}{d\tau}\frac{dx^\nu}{d\tau}=\frac{1}{2}\mu g_{\mu\nu}{\dot{x}}^\mu{\dot{x}}^\nu.
\end{equation}

Assuming that the test particle is incident along the equatorial plane, there is no motion in the $\theta$ direction, i.e. $\frac{d\theta}{d\tau}=0$. Therefore, the momentum of the test particle in the $\theta$ direction is zero, given by
\begin{equation}\label{11}
p_\theta=\frac{\partial L}{\partial\dot{\theta}}=\mu g_{22}\dot{\theta}=0.
\end{equation}
The angular momentum $\delta J$ and energy $\delta E$ of a moving particle in the spacetime of rotating black-bounce can be expressed as
\begin{equation}\label{12}
\delta J=\frac{\partial L}{\partial\phi}=\mu g_{3\nu}{\dot{x}}^\nu,
\end{equation}
\begin{equation}\label{13}
\delta E=-\frac{\partial L}{\partial\dot{t}}=-\mu g_{0\nu}{\dot{x}}^\nu.
\end{equation}
The change in angular momentum $J'$ and energy  $M'$  of the black hole itself after capturing a particle is given by

\begin{equation}\label{14}
\begin{split}
M\to M'=M+\delta E,\\
J\to J'=J+\delta J.
\end{split}
\end{equation}

In order to investigate whether the event horizon of this spacetime can be violated, it is necessary to first calculate the conditions under which the test particle can enter the event horizon. This is because if the angular momentum of the test particle is too large, the centrifugal repulsion may cause the test particle to escape the capture by spacetime and thus fail to reach the interior of the event horizon. Then, the conditions for the composite system formed by particles crossing the event horizon to potentially disrupt the event horizon of this spacetime need to be calculated. Only when both of these conditions are satisfied can the structure inside the event horizon be exposed to external observers.

For a massive particle, the time component of the four-velocity that describes its motion is non-zero. The four-velocity must be less than the speed of light and therefore must be a timelike unit vector, denoted as
\begin{equation}\label{15}
U^\mu U_\mu=g_{\mu\nu}\frac{dx^\mu}{d\tau}\frac{dx^\nu}{d\tau}=\frac{1}{\mu ^2}g^{\mu\nu}P_\mu P_\nu=-1.
\end{equation}
Substituting equations \eqref{11},\eqref{12} and \eqref{13} into equation \eqref{15} yield
\begin{equation}\label{16}
g^{00}\delta E^2-2g^{03}\delta E\delta J+g^{11}p^2_r+g^{33}\delta J^2=-\mu ^2.
\end{equation}
To solve equation \eqref{16} for the energy $\delta E$ of the test particle
\begin{equation}\label{17}
\begin{split}
\delta\ E=&\frac{g^{03}}{g^{00}}\delta\ J\\
&\pm\frac{1}{g^{00}}\sqrt{\left[{(g^{03})}^2{\delta J}^2-g^{00}(g^{33}{\delta J}^2+g^{11}P_r^2+\mu ^2\right]}.
\end{split}
\end{equation}
The test particle moves along the equatorial plane towards the event horizon, and in this case, the geodesic of its motion is timelike and future-directed. The conditions that need to be satisfied are
\begin{equation}\label{18}
\frac{dt}{d\tau}>0.
\end{equation}
Expand and solve equations \eqref{12} and \eqref{13} to obtain
\begin{equation}\label{19}
\dot{t}=\frac{dt}{d\tau}=-\frac{g_{33}\delta E+g_{03}\delta J}{\mu (g_{00}g_{33}-{g_{03}}^2)}.
\end{equation}
Because the condition is $\frac{dt}{d\tau}>0$, its energy must be satisfied as
\begin{equation}\label{20}
\delta E>-\frac{g_{03}}{g_{33}}\delta J.
\end{equation}
From the condition given by equation \eqref{20}, we can deduce that the energy in equation \eqref{17} can only take a negative sign, which means
\begin{equation}\label{21}
\begin{split}
\delta E=&\frac{g^{03}}{g^{00}}\delta J-\\
&\frac{1}{g^{00}}\sqrt{\left[{(g^{03})}^2{\delta J}^2-g^{00}(g^{33}{\delta J}^2+g^{11}P_r^2+\mu ^2\right]}.
\end{split}
\end{equation}
In order for test particles to fall into the interior of this event horizon, the condition that the equation \eqref{20} points in the future becomes
\begin{equation}\label{22}
\delta J<-\lim_{r\to r_h}{\frac{g_{33}}{g_{03}}}\delta E.
\end{equation}
Substituting equation (6) into equation \eqref{22} yields
\begin{equation}\label{23}
\delta J<-\lim_{r\to r_h}{\frac{g_{33}}{g_{03}}}\delta E=\frac{r_h^2+m^2+a^2}{a}\delta E.
\end{equation}

On one hand, if the energy and angular momentum relationship of the test particle is such that $\delta J \gg \delta E$, there would be a centrifugal repulsive force that is larger than the attractive gravitational force. This would cause the particle to miss the collision point without colliding. If this happens, in this spacetime, the particle would have no possibility of entering the interior of the event horizon, which is not the desired outcome. Therefore, there should be an upper limit on the angular momentum $\delta J$ of the test particle. If $\delta J$ exceeds this limit, the test particle cannot enter the interior of the event horizon. Thus, from equation \eqref{23} above, we can determine that the upper limit for the angular momentum $\delta J$ of the test particle is
\begin{equation}\label{24}
\delta J_{max}<\frac{\delta E}{\Omega_H}=\frac{r_h^2+m^2+a^2}{a}\delta E.
\end{equation}

On the other hand, in order to overspin the event horizon of a rotating Black-Bounce black hole, the angular momentum $\delta J$  and energy $\delta E$ of the test particle must satisfy the conditions specified by the event horizon equation. That is, the composite system formed after the black hole absorbs the energy and angular momentum of the particle satisfies the following conditions
\begin{equation}\label{25}
J'>{M'}^2\sqrt{\beta^2-\frac{m^2}{M^2}}.
\end{equation}
To facilitate the subsequent analysis, let the parameter be denoted as 
\begin{equation}\label{26}
\alpha=\sqrt{\beta^2-\frac{m^2}{M^2}}.
\end{equation}
At this time, the conditions that the angular momentum $\delta J$ and energy $\delta E$ must satisfy become the following $J'>\alpha M'^2$.

To satisfy the composite system condition $J'>\alpha M'^2$ mentioned above, substitute equation \eqref{14} into the composite system condition, resulting in
\begin{equation}\label{27}
J+\delta J>\alpha(M+\delta E)^2,
\end{equation}
expanding this expression yield
\begin{equation}\label{28}
\begin{split}
\delta J>\alpha\delta E^2+2\alpha M\delta E+\left({\alpha M}^2-J\right).
\end{split}
\end{equation}
Clearly, from equation \eqref{28} can be seen that in order to disrupt the event horizon of a rotating black-bounce black hole, the condition that must be satisfied by the test particle's angular momentum $\delta J$ is as follows
\begin{equation}\label{29}
\begin{split}
\delta J_{min}>\left({\alpha M}^2-J\right)+\alpha\delta E^2+2\alpha M\delta E.
\end{split}
\end{equation}
The above equation represents the minimum angular momentum required for a test particle to disrupt the event horizon of this spacetime.

Therefore, only when the angular momentum of the test particle satisfies the conditions given by equations \eqref{24} and \eqref{29}, will the event horizon of this rotating Black-Bounce hole be disrupted, revealing its internal structure to external observers.

For a rotating Black-Bounce black hole, we only consider two cases: extreme and near-extreme. For the extreme case, that is $a=M\alpha$. Then the event horizon of the black hole is 
\begin{equation}\label{30}
r_h=\Gamma=M\beta.
\end{equation}
If the analysis only considers the first-order approximation of energy $\delta E$, then the condition for disrupting the event horizon of this spacetime becomes
\begin{equation}\label{31}
\delta J_{max}<\frac{\delta E}{\Omega_H}=\frac{r_h^2+m^2+a^2}{a}\delta E,
\end{equation}
\begin{equation}\label{32}
\delta J_{min}>2\alpha M\delta E.
\end{equation}
Combining equations \eqref{30} and \eqref{31}, we obtain the following calculation results
\begin{equation}\label{33}
\delta J_{max}<\left(2M\alpha+\frac{{2m}^2}{M\alpha}\right)\delta E.
\end{equation}

From equations \eqref{32} and \eqref{33} above  can be visually seen that the angular momentum $\delta J$ of the test particle can satisfy both conditions simultaneously. In the extreme case, the event horizon of a rotating Black-Bounce hole can be disrupted. Only when the parameter $m=0$, the event horizon of this extreme spacetime model cannot be destroyed. When $m=k=0$ and $n=1$, this spacetime becomes a Kerr black hole. When the aforementioned value is taken in the equation, the angular momentum of the test particle cannot satisfy both conditions simultaneously, and in this case, the event horizon of this spacetime cannot be disrupted. This scenario is consistent with the conclusion in general relativity that the event horizon of an extreme Kerr black hole cannot be disrupted by test particles\cite{Jacobson2009OverspinningAB}.

 The above analysis only considers the first-order approximation. If we take into account the second-order small quantities, according to equation (\ref{29}), although the second-order small quantity $\alpha\delta E^2$ increases the lower limit of angular momentum for the test particle. However, we found that when only considering first-order approximation, our conclusion is that as long as parameter $m$ exists, equations (\ref{32}) and (\ref{33}) can be satisfied simultaneously. In other words, under the first-order approximation, we have already obtained the possibility of destroying the event horizon. Therefore, when considering the second-order small quantities, their influence on the result is minimal, and the possibility of event horizon disruption still exists. Therefore, in the subsequent analysis, we will also consider only the first-order approximation. Only when the first-order approximation is inconclusive, we will incorporate the analysis of the second-order small quantities.

For another case, which corresponds to a near-extreme situation, that is, $a\sim\alpha M$. Here, we also consider only the first-order approximation of the energy $\delta E$ for the test particle. In this case, the conditions for the test particle to pass through the event horizon and the conditions for disrupting the event horizon become
\begin{equation}\label{34}
\delta J_{max}<\frac{\delta E}{\Omega_H}=\frac{r_h^2+m^2+a^2}{a}\delta E,
\end{equation}
\begin{equation}\label{35}
\delta J_{min}>2\alpha M\delta E+\left(\alpha M^2-J\right).
\end{equation}
For the expression $a\sim\alpha M$ mentioned above, a dimensionless parameter $\epsilon$ can be defined to describe the degree of approaching the extreme. It is defined as the following
\begin{equation}\label{36}
\frac{a^2+m^2}{\Gamma^2}=1-\epsilon^2.
\end{equation}
The parameter $\epsilon$ is a number approaching zero, that is $\epsilon\ll1$. When $\epsilon=0$, the equation becomes the extreme case. From equations \eqref{34} and \eqref{35}, it can be concluded that in order to disrupt the event horizon of this spacetime in the near-extreme scenario while considering only the first-order approximation, the term $(\alpha M^2-J)$ can be neglected as a second-order small quantity. Therefore, by analyzing equations (\ref{34}) and (\ref{35}), we obtain
\begin{equation}\label{37}
\frac{1}{\Omega_H}-2\alpha M>0.
\end{equation}
If the subsequent calculation results satisfy the aforementioned equation, then the event horizon of this black hold can be disrupted in the near-extreme scenario. By combining equations \eqref{7} and \eqref{37}, we can calculate
\begin{equation}\label{38}
r_h=\left(1+\epsilon\right)\Gamma=\left(1+\epsilon\right)\beta M.
\end{equation}
Combining equations (\ref{6}) , (\ref{37})and (\ref{38}), the result is
\begin{equation}\label{39}
\begin{split}
&\frac{1}{\Omega_H}-2\alpha M=\\
&\frac{{2m}^2+2\left(M^2\alpha^2+m^2\right)\epsilon+\left(M^2\alpha^2-m^2\right)\epsilon^2-{2M}^2\alpha^2O\left(\epsilon^4\right)}{\sqrt{\left(1-\epsilon^2\right)\left(M^2\alpha^2+m^2\right)-m^2}}.
\end{split}
\end{equation}

Based on equation \eqref{39},it can be easily deduced that the disruption of the event horizon of a rotating Black-Bounce black hole depends on the value of parameter $m$. In this equation, parameter $m$ is a non-negative value, so in the near-extreme scenario, the event horizon of the entire spacetime can be easily disrupted. For the case of $m=k=0$ and $n=1$, this spacetime degenerates into a Kerr black hole. An analysis of the equation reveals that in this case, the event horizon can be disrupted. This conclusion aligns with the notion that an approximately extreme Kerr black hole can expose its singularity by testing particles to disrupt the horizon\cite{Jacobson2009OverspinningAB}.

\section{\label{sec:level4}Destruction of Event Horizon by Scalar Fields}
Another way to disrupt the event horizon of a non-singular rotating black hole is by scattering a scalar field with large angular momentum onto a near-extreme or extreme black hole. This idea was proposed by Semiz in 2011 and has since been further developed and refined by researchers like Gwak\cite{PhysRevD.88.064043,PhysRevD.92.104021}.

Therefore, in this section, we will examine the weak cosmic censorship conjecture for rotating Black-Bounce black hole by referencing the ideas of previous researchers. Following the approach of previous studies, we will scatter a scalar field with large angular momentum onto this spacetime and investigate the possibility of disrupting the event horizon of the rotating Black-Bounce black hole in extreme and near-extreme scenarios. Similar to prior research, we will utilize a scalar field with large angular momentum to scatter onto this spacetime and study the cases where the disruption of the event horizon of the rotating Black-Bounce black hole occurs in extreme and near-extreme conditions.
\subsection{\label{sec:level4.1}Scattering of Massive Scalar Fields}
Scattering occurs when a scalar field is incident on a rotating Black-Bounce black hole. Assuming the mass of this scalar field $\psi$ is given by $\mu$, the motion equation of this scalar field can be described using the Klein-Gordon equation, which is given by
\begin{equation}\label{40}
\nabla_\mu\nabla^\nu-\mu^2\psi=0.
\end{equation}
According to the definitions of covariance and contravariance, the equation can be written as
\begin{equation}\label{41}
\frac{1}{\sqrt{-g}}\partial_\mu\left(\sqrt{-g}g^{\mu\nu}\partial_\nu\psi\right)-\mu^2\psi=0.
\end{equation}
The determinant of the metric for a rotating Black-Bounce black hole is
\begin{equation}\label{42}
g=-\Sigma^4\sin^2{\theta}.
\end{equation}
The contravariant metric tensor is
\begin{equation}\label{43}
g^{\mu\nu}=\frac{\Delta^{\mu\nu}}{g}.
\end{equation}
By substituting equations \eqref{42} and \eqref{43} into equation \eqref{41}, the result can be calculated as
\begin{equation}\label{44}
\begin{split}
-\frac{\left(r^2+m^2+a^2\right)^2-a^2\Delta\sin^2{\theta}}{\Delta\Sigma^2}\frac{\partial^2\psi}{\partial t^2}-\frac{4aM\beta}{\Delta\Sigma^2}\frac{\partial^2\psi}{\partial t\partial\phi}\\
+\frac{1}{\Sigma^2}\frac{\partial}{\partial r}\left(\Delta\frac{\partial\psi}{\partial r}\right)+\\
\frac{1}{\Sigma^2\sin^2\theta}\frac{\partial}{\partial\theta}\left(\sin\theta\frac{\partial\psi}{\partial\theta}\right)+\frac{\Delta-a^2\sin^2\theta}{\Delta\Sigma^2}\frac{\partial^2\psi}{\partial\phi^2}-\mu^2\phi=0.
\end{split}
\end{equation}
In order to separate the variables in the equation above, we can assume the solution for the scalar field $\psi$ to have the following form
\begin{equation}\label{45}
\psi\left(t,r,\theta,\phi\right)=e^{-i\omega t}R\left(r\right)S_{lm_0}(\theta)e^{im_0\phi}.
\end{equation}
The term $S_{lm_0}(\theta)$ in the equation represents the angular spherical harmonic function, where $l$ and $m_0$ are constants for angular separation of variables, taking positive integer values. By substituting equation (\ref{45}) into the scalar field equation (\ref{44}), we obtain the following form
\begin{widetext}
\begin{equation}\label{46}
\begin{split}
&\left[\frac{1}{\sin^2{\theta}}\frac{d}{d\theta}\left(\sin{\theta}\frac{dS_{lm_0}\left(\theta\right)}{d\theta}\right)-\left(a^2\omega^2\sin^2{\theta}+\frac{{m_0}^2}{\sin^2{\theta}}+\mu^2a^2\cos^2{\theta}\right)S_{lm_0}\left(\theta\right)\right]R\left(r\right)+\\
&\left[\frac{d}{dr}\left(\Delta\frac{dR}{dr}\right)+\left(\frac{(r^2+m^2+a^2)^2}{\Delta}\omega^2-\frac{4aM\beta}{\Delta}m_0\omega+\frac{m_0^2a^2}{\Delta}-\mu^2(r^2+m^2)\right)R(r)\right]S_{lm_0}(\theta)=0.
\end{split}
\end{equation}
\end{widetext}
By separating variables in equation (\ref{46}), we obtain the angular equation for the scalar field as follows
\begin{widetext}
\begin{equation}\label{47}
\frac{1}{\sin^2{\theta}}\frac{d}{d\theta}\left(\sin{\theta}\frac{dS_{lm_0}\left(\theta\right)}{d\theta}\right)-\left(a^2\omega^2\sin^2{\theta}+\frac{{m_0}^2}{\sin^2{\theta}}+\mu^2a^2\cos^2{\theta}-\lambda_{lm_0}\right)S_{lm_o}\left(\theta\right)=0,
\end{equation}
\end{widetext}
and the radial equation for the scalar field as follows

\begin{equation}\label{48}
\begin{split}
\frac{d}{dr}\left(\Delta\frac{dR}{dr}\right)+\left(\frac{(r^2+m^2+a^2)^2}{\Delta}\omega^2-\frac{4aM\beta}{\Delta}m_0\omega \right.\\
\left.+\frac{m_0^2a^2}{\Delta}-\mu^2(r^2+m^2)-\lambda_{lm_0}\right)R(r)=0.
\end{split}
\end{equation}

In the above two equations $\lambda_{lm_0}$ is a separation constant and represents the eigenvalue of the angular spherical harmonic function. By solving equation \eqref{47}, it is found that its solution is a spherical function. Due to the normalization of spherical functions, when calculating the energy flux in the subsequent steps by integrating over the entire event horizon, the integral of the spherical function is equal to one. Therefore, we focus more on finding the radial solution of the scalar field equation. For convenience in solving the equation, we introduce the tortoise coordinate $r_\ast$ and define the tortoise coordinate as
\begin{equation}\label{49}
\frac{dr}{dr_\ast}=\frac{\Delta}{r^2+m^2+a^2}.
\end{equation}
After introducing the tortoise coordinate $r_\ast$, it can cover the entire region outside the event horizon of a rotating Black-Bounce black hole. Substituting the tortoise coordinate into the radial equation (\ref{48}) for the scalar field and solving it, we obtain
\begin{equation}\label{50}
\begin{split}
&\frac{\Delta}{\left(r^2+a^2+m^2\right)^2}\frac{d}{dr}\left(r^2\right)\frac{dR}{dr_\ast}+\frac{d^2R}{d{r_\ast}^2}+\\
&\left[\left(\omega-\frac{m_0a}{r^2+a^2+m^2}\right)^2+\frac{2\Delta am_0\omega}{\left(r^2+a^2+m^2\right)^2}\right.\\
&\left.-\frac{\Delta}{\left(r^2+a^2+m^2\right)^2}\left(\mu^2\left(r^2+m^2\right)+\lambda_{lm_0}\right)\right]R=0.
\end{split}
\end{equation}

In the subsequent analysis, we are primarily concerned with the energy flux and angular momentum flux of the scalar field incident on the event horizon of a rotating Black-Bounce black hole. Therefore, in the vicinity of the horizon, we make the following approximation $r\cong r_h$, which means
\begin{equation}\label{51}
\Delta\cong0.
\end{equation}
Substituting equation (\ref{51}) into equation (\ref{50}) yields the following approximation
\begin{equation}\label{52}
\frac{d^2R}{d{r_\ast}^2}+\left(\omega-\frac{m_0a}{r_h^2+a^2+m^2}\right)^2R=0.
\end{equation}
By substituting equation (\ref{6}) into equation (\ref{52}), the radial equation for the scalar field can be expressed as
\begin{equation}\label{53}
\frac{d^2R}{d{r_\ast}^2}+\left(\omega-m_0\Omega_H\right)^2R=0.
\end{equation}
The solution can be expressed in exponential form as
\begin{equation}\label{54}
R(r)\sim exp\left[\pm i(\omega-m_0\Omega_H)r_\ast\right].
\end{equation}
In the solution above, the positive and negative signs correspond to the outgoing and ingoing waves, respectively. We are primarily interested in the wave modes absorbed by a rotating black-bounce balck hole.  Therefore, we consider the negative sign, which corresponds to the ingoing wave and is more appropriate for our analysis. Hence, the solution to the radial equation for the scalar field is
\begin{equation}\label{55}
R\left(r\right)=exp\left[-i(\omega-m_0\Omega_H)r_\ast\right].
\end{equation}
 The approximate solution for the scalar field near the event horizon is
\begin{equation}\label{56}
\psi\left(t,r,\theta,\phi\right)=exp\left[-i(\omega-m_0\Omega_H)r_\ast\right]e^{-i\omega t}S_{lm_0}(\theta)e^{im_0\phi}.
\end{equation}
With this approximate solution, we can now calculate the energy flux and angular momentum flux of the scalar field near the event horizon. This enables us to investigate whether a rotating Black-Bounce black hole can desytroy the event horizon by absorbing energy from the scalar field.

Let's assume that a scalar field with mode $(l, m_0)$ is incident on this spacetime. During the process of incidence, the spacetime absorbs a fraction of the energy from the scalar field and reflects another fraction of the energy. The absorbed portion of the scalar field's energy will be converted into angular momentum and energy of the composite system. By calculating the fluxes of energy and angular momentum of the scalar field around the event horizon, we can obtain the corresponding values of angular momentum and energy.

The stress-energy tensor of a scalar field $(\psi)$ with mass $(\mu)$ can be expressed in the following form
\begin{equation}\label{57}
T_{\mu\nu}=\partial_\mu\psi\partial_\nu\psi^\ast-\frac{1}{2}g_{\mu\nu}\left(\partial_\mu\psi\partial^\nu\psi^\ast+\mu^2\psi\psi^\ast\right).
\end{equation}
By substituting equations (\ref{1}) and (\ref{56}) into equation (\ref{57}), we obtain
\begin{equation}\label{58}
\begin{split}
T_t^r=&\frac{(r_h^2+a^2+m^2)}{\Sigma^2}\\
&\omega(\omega-m_0\Omega_H)S^\ast_{lm_0}(\theta)e^{im_0\phi}S^\ast_{l'm'_0}(\theta)e^{-im_0\phi},
\end{split}
\end{equation}
and
\begin{equation}\label{59}
\begin{split}
T_\phi^r=&\frac{(r_h^2+a^2+m^2)}{\Sigma^2}\\
&m_0\left(\omega-m_0\Omega_H\right)S^\ast_{lm_0}(\theta)e^{im_0\phi}S\ast_{l'm'_0}(\theta)e^{-im_0\phi}.
\end{split}
\end{equation}
Therefore, the energy flux through the event horizon is
\begin{equation}\label{60}
\frac{dE}{dt}=\iint{T_t^r\sqrt{-g}}d\theta d\phi=\omega(\omega-m_0\Omega_H)\left[{r_h}^2+a^2+m^2\right],
\end{equation}
and the angular momentum flux through the event horizon is
\begin{equation}\label{61}
\frac{dJ}{dt}=\iint{T_\phi^r\sqrt{-g}}d\theta d\phi=m_0\left(\omega-m_0\Omega_H\right)\left[{r_h}^2+a^2+m^2\right].
\end{equation}

From the above two equations, where $g$ is the determinant of the metric, it can be easily deduced that when $\omega > m_0\Omega_H$, the values of angular momentum and energy flux through the horizon are positive. This means that, in this scenario, a rotating Black-Bounce black hole gains energy $dE$ and angular momentum $dJ$ from the scalar field. On the contrary, when $\omega < m_0\Omega_H$, the values of angular momentum and energy flux through the horizon are negative. This implies that, at this moment, the spacetime does not gain energy from the scalar field. Instead, the scalar field extracts the corresponding energy from the black hole. This phenomenon is known as black hole superradiance.

For a very small time interval $dt$, the changes in angular momentum and mass of this spacetime are
\begin{equation}\label{62}
dE=\omega(\omega-m_0\Omega_H)\left[{r_h}^2+a^2+m^2\right]dt,
\end{equation}
and
\begin{equation}\label{63}
dJ=m_0(\omega-m_0\Omega_H)\left[{r_h}^2+a^2+m^2\right]dt.
\end{equation}

Through equations (\ref{62}) and (\ref{63}), we can determine the energy and angular momentum of a massive scalar field that passes through the event horizon. In other words, these equations provide the values of absorbed energy and angular momentum from the scalar field in this spacetime model. By analyzing the absorbed angular momentum and energy, we can assess whether the event horizon of this spacetime can be destroyed in extreme and near-extreme scenarios, and whether observers outside can observe the internal structure of the event horizon.
\subsection{\label{sec:level4.2}Destruction of Event Horizon of a Rotating Black-Bounce Black Hole after Scalar Field Scattering}
In this section, we main focus is to discuss the destruction of the event horizon by a scalar field with large angular momentum scattering off arotating Black-Bounce black hole. By considering the scattering of a monochromatic scalar field with a frequency of $\omega$ and an angular quantum number of $m_0$ onto the rotating Black-Bounce black hole, we examine whether this scalar field can destroy the event horizon of the spacetime. Additionally, we explore the influence of the parameter $m$ on the destruction of the horizon.

For a continuous process of scalar field scattering, the calculus approach is employed by breaking down the continuous process into numerous infinitesimal time intervals $dt$. Then, each time interval $dt$ is analyzed separately, and the only difference between the analysis of each interval is the different initial state parameters of the black hole. The analysis process remains the same for each interval.

According to section \ref{sec:level2}, it is known that the event horizon of the spacetime can be destroyed when the black hole spin parameter $a$ changes. Using the same approach, before the incident of a scalar field, the initial mass of the black hold is $M$ and the initial angular momentum is $J$. After the incidence of the scalar field, the mass of the the black hole becomes $M' = M + dE$, and the angular momentum becomes $J' = J + dJ$. For the composite system formed by a black hole after the incidence of a scalar field,by using the formula for the event horizon, we can determine whether the event horizon of the spacetime is destroyed. We only need to consider the sign of the expression $\alpha M'^2 - J'$. If the sign is positive, then the event horizon of the spacetime cannot be destroyed, and the event horizon will continue to exist. If the sign is negative, then the event horizon of the spacetime is destroyed, thereby exposing the internal structure of the spacetime to observers at infinity.

As mentioned above, Our analysis focuses solely on the variation of the composite system within the extremely short time interval dt. Within this time interval dt, the rotating Black-Bounce black hole absorbs energy dE and angular momentum dJ from the scalar field, resulting in the total energy and angular momentum of the composite system being
\begin{equation}\label{64}
\begin{split}
\alpha M'^2-J'=\left(\sigma M^2-J\right)+2\sigma MdE+\sigma{dE}^2-dJ.
\end{split}
\end{equation}
When considering only the first-order terms, the above equation can be simplified as
\begin{equation}\label{65}
\alpha{M'}^2-J'=\left(\alpha M^2-J\right)+2\alpha MdE-dJ.
\end{equation}
Substituting the energy and angular momentum absorbed by the spacetime from the scalar field, as analyzed in section \ref{sec:level4.1}, into equations (\ref{62}) and (\ref{63}), and then plugging them into equation (\ref{65}), we have
\begin{equation}\label{66}
\begin{split}
\alpha{M'}^2-J'=&\left(\alpha M^2-J\right)+2\alpha M{m_0}^2\left(\frac{\omega}{m_0}-\mathrm{\Omega}_H\right)\\
&\left(\frac{\omega}{m_0}-\frac{1}{2\alpha M}\right)\left({r_h}^2+a^2+m^2\right)dt.
\end{split}
\end{equation}

For a rotating Black-Bounce black hole in the extreme case, where $\alpha M^2 = J$, equation (\ref{66}) becomes
\begin{equation}\label{67}
\begin{split}
\alpha{M'}^2-J'=&2\alpha M{m_0}^2\left(\frac{\omega}{m_0}-\mathrm{\Omega}_H\right)\\&\left(\frac{\omega}{m_0}-\frac{1}{2\alpha M}\right)\left({r_h}^2+a^2+m^2\right)dt.
\end{split}
\end{equation}
From the angular velocity formula (\ref{6}), in the extreme case, the angular velocity of a rotating black-bounce black hole can be simplified as
\begin{equation}\label{68}
\Omega_H=\frac{a}{r_h^2+a^2+m^2}=\frac{1}{2\alpha M+\frac{2m^2}{\alpha M}}\le\frac{1}{2\alpha M}.
\end{equation}
In the above equation, equality holds only when the parameter $m$ equals zero. Through analysis of the equation, it can be observed that the presence of the parameter $m$ causes a shift in the angular velocity at the event horizon of the spacetime. This result implies that the event horizon of the spacetime model may be disrupted.

In the subsequent analysis, by examining equation (\ref{67}), it can be readily concluded that assuming an incident scalar field is injected with the following mode
\begin{equation}\label{69}
\frac{\omega}{m_0}=\frac{1}{2}\left(\frac{1}{2\alpha M}+\mathrm{\Omega}_H\right).
\end{equation}
The state of the composite system transforms into
\begin{equation}\label{70}
\begin{split}
\alpha{M'}^2&-J'\\
&=\frac{1}{2}\alpha M{m_0}^2\left(\frac{1}{2\alpha M}-\mathrm{\Omega}_H\right)\\
&\left(\mathrm{\Omega}_H-\frac{1}{2\alpha M}\right)\left({r_h}^2+a^2+m^2\right)dt\\
&=-\frac{1}{2}\alpha M{m_0}^2\left(\frac{1}{2\alpha M}-\mathrm{\Omega}_H\right)^2\left({r_h}^2+a^2+m^2\right)dt.
\end{split}
\end{equation}
According to equation (\ref{70}),it can be easily derived that
\begin{equation}\label{71}
{M'}^2\alpha-J'\le0.
\end{equation}

The equality in the equation holds only when the parameter $m$ equals zero. In this case, the event horizon of the spacetime cannot be destroyed. When $m=k=0$ and $n=1$, this spacetime becomes a Kerr black hole, which analysis shows cannot be destroyed either, consistent with the fact that extremal Kerr black holes cannot be destroyed by scalar fields in general relativity. However, when $m$ takes certain values, the event horizon of this rotating black-bounce black hole can be destroyed. Specifically, when $m$ takes other values, equation (\ref{71}) is always less than zero. Clearly, after absorbing the scalar field modes described above, this rotating spacetime lacks an event horizon. In other words, this scalar field mode can destroy the event horizon of the spacetime, revealing the internal structure of the event horizon.

Actually, scalar field modes that can destroy the event horizon are not limited to the mode described above. There are other modes as well that can also destroy the event horizon. In other words, the scalar field modes exist within a range, and all modes within this range can destroy the event horizon. By combining equations (\ref{67}) and (\ref{68}), we can obtain that the scalar field mode satisfies the following form
\begin{equation}\label{72}
\frac{1}{2\alpha M+\frac{{2m}^2}{\alpha M}}<\frac{\omega}{m_0}<\frac{1}{2\alpha M}.
\end{equation}
If the scalar field mode satisfies the conditions given by the above equation, then the composite system will necessarily satisfy the following form
\begin{equation}\label{73}
{M'}^2\alpha-J'<0.
\end{equation}
In other words, when the range of scalar field modes satisfies equation (\ref{72}), the event horizon of an extreme rotating Black-Bounce black hole can be disrupted. Moreover, it can be obviously observed from this equation that a large range of scalar field modes is allowed to satisfy it when the parameter $m$ is larger. When the parameter $m=0$, the event horizon cannot be destroyed.

For the near-extremal case, that is, when $\alpha M^2 \sim J$, the expression is as follows
\begin{equation}\label{74}
\begin{split}
{\alpha M'}^2-J'=&\left(\alpha M^2-J\right)\\
&+2\alpha M{m_0}^2\left(\frac{\omega}{m_0}-\mathrm{\Omega}_H\right)\\
&\left(\frac{\omega}{m_0}-\frac{1}{2\alpha M}\right)\left({r_h}^2+a^2+m^2\right)dt.
\end{split}
\end{equation}
Analyzing equation (\ref{74}), we assuming the mode of the incident scalar field is 
\begin{equation}\label{75}
\frac{\omega}{m_0}=\frac{1}{2}\left(\frac{1}{2\alpha M}+\mathrm{\Omega}_H\right).
\end{equation}
In this case, equation (\ref{74}) becomes
\begin{equation}\label{76}
\begin{split}
{\alpha M'}^2-J'=&\left(\alpha M^2-J\right)-\frac{1}{8\alpha M}\mathrm{\Omega}_H{m_0}^2\\&\left(\frac{1}{\mathrm{\Omega}_H}-2\alpha M\right)^2\left({r_h}^2+a^2+m^2\right)dt.
\end{split}
\end{equation}

Similar to the analysis of considering test particles incident on a rotating Black-Bounce black hole in the near-extremal case as before, we can define an infinitesimal and dimensionless parameter $\epsilon$ to represent the deviation of the near-extremal case from the extremal case. It can be expressed using the following equation
\begin{equation}\label{77}
\frac{a^2+m^2}{M^2\beta^2}=1-\epsilon^2.
\end{equation}
Since  $\epsilon$ is a small parameter approaching zero, we can perform a Taylor expansion of equation (\ref{77}) and substitute it into equation (\ref{76}). This yields
\begin{widetext}
\begin{equation}\label{78}
\begin{split}
{{\alpha M}'}^2-J'=&\left(\alpha M^2-J\right)-\frac{1}{8\alpha M}\mathrm{\Omega}_H{m_0}^2\left(\frac{1}{\mathrm{\Omega}_H}-2\alpha M\right)^2\left({r_h}^2+a^2+m^2\right)dt\\
=&\left[\frac{1}{2}\frac{\left(M^2\alpha^2+m^2\right)}{\sqrt{M^2\alpha^2-\left(M^2\alpha^2+m^2\right)\epsilon^2}}\epsilon^2-\alpha M^2O\left(\epsilon^4\right)\right]\\
&-\frac{1}{8\alpha M}\mathrm{\Omega}_H{m_0}^2\left(\frac{{2m}^2+2\left(M^2\alpha^2+m^2\right)\epsilon+\left(M^2\alpha^2-m^2\right)\epsilon^2-{2M}^2\alpha^2O(\epsilon^4)}{\sqrt{M^2\alpha^2-\left(M^2\alpha^2+m^2\right)\epsilon^2}}\right)^2\left({r_h}^2+a^2+m^2\right)dt.
\end{split}
\end{equation}
\end{widetext}
The process we are analyzing is a composite state within an extremely short time interval $dt$ , which means that both $dt$ and $\epsilon$ here are considered to be first-order small quantities. Through the analysis of the above equation, we can conclude that only when $m=0$, we have $\alpha M'^2-J'>0$. In this case, the event horizon of a near-extremal rotating Black-Bounce black hole cannot be destroyed. Furthermore, when $m=k=0$ and $n=1$, the spacetime becomes a Kerr black hole, and the aforementioned analysis is consistent with the fact that a near-extremal Kerr black hole cannot be destroyed by a scalar field. However, when the parameter $m\neq 0$, it is evident from equation (\ref{78}) that
\begin{equation}\label{79}
\alpha M'^2-J'<0.
\end{equation}
This indicates that the event horizon of a near-extremal rotating Black-Bounce black hole can be destroyed. In other words, whether the event horizon of this spacetime can be destroyed by a scalar field in the near-extremal case depends on the choice of the parameter $m$. It can only be destroyed when the parameter $m\neq 0$.

In summary, we have discussed the destruction of the event horizon of rotating black-bounce black holes under the incidence of a scalar field in the extreme and near-extreme cases. It is found that regardless of whether it is in the extreme or near-extreme case, the destruction of the event horizon of the rotating black-bounce black hole depends on the value of the parameter $m$. As long as the value of $m\neq 0$, the event horizon of the rotating Black-Bounce black hole can be destroyed in both extreme and near-extreme cases.
\section{Discussion and conclusions}
In this paper, we primarily examine the weak cosmic censorship conjecture by throwing test particles and using a scalar field with large angular momentum into a rotating Black-Bounce black hole. Through computational analysis, we have discovered that the event horizon of this spacetime is potentially destructible when test particles are thrown into the rotating Black-Bounce black hole under extreme conditions. However, the event horizon of this extreme spacetime model cannot be destroyed when the parameter $m=0$. Additionally, as the parameter $m$ increases, it becomes easier to destroy the event horizon, and the range of angular momentum values that lead to the destruction of the event horizon also increases. This observation may suggest an inherent connection between the physical origin of Black-Bounce black holes and the absorption of particles by black holes, thereby indicating the possible existence of Black-Bounce spacetime in the actual universe. Furthermore, in the near-extreme conditions, our analysis reveals that this spacetime model is also highly susceptible to destruction. Unlike the extreme case, the event horizon of the rotating Black-Bounce black hole can be destroyed regardless of the value of the parameter $m$. Moreover, the larger the value of the parameter $m$, the more easily the event horizon can be destroyed.

When we use a scalar field with large angular momentum for scattering, we find that regardless of whether it is in extreme or near-extreme conditions, the event horizon of this spacetime can be destroyed. Moreover, in the extreme case, we observe that the larger the value of the parameter $m$, the greater the range of scalar field modes that can destroy the event horizon of the rotating Black-Bounce black hole. In other words, the event horizon becomes more susceptible to destruction as the parameter $m$ increases.

Regardless of whether it is the destruction of event horizon caused by test particles or scalar fields, when $m=k=0$, $n=1$, the spacetime becomes a Kerr black hole. the analysis results in this case are consistent with the research findings of others regarding Kerr black holes.

Through analysis of this spacetime model, we have found that the event horizon of the rotating Black-Bounce black hole is easily destroyed, which seems to contradict the weak cosmic censorship conjecture and opens up possibilities for exploring the internal structure of the event horizon. However, since our metric for the rotating Black-Bounce black hole is obtained under specific conditions, extensive research is needed to determine whether the violation of the weak cosmic censorship conjecture is a universal occurrence in this spacetime.

Our work reveals that the physical origin of Black-Bounce black holes may be related to the absorption of particles by black holes, providing new avenues for further understanding of the Black-Bounce spacetime. In future research, we will explore the connection between large-mass incident black holes and the formation of Black-Bounce black holes.
\section{acknowledgements}
We acknowledge the anonymous referee for a constructive report that has significantly improved this paper. We acknowledge the  Special Natural Science Fund of Guizhou University (grant
No. X2020068 and No. X2022133) and the financial support from the China Postdoctoral Science Foundation funded project under grants No. 2019M650846.

\nocite{*}
\bibliographystyle{unsrt}
\bibliography{WCCC}

\end{document}